\documentstyle[twocolumn,aps]{revtex}
\begin{document}
\newcommand{\beq}{\begin{equation}}
\newcommand{\eeq}{\end{equation}}

\title{Master equation approach to protein folding and kinetic traps}
\author{
Marek Cieplak$^1$, Malte Henkel$^2$, Jan Karbowski$^3$,
and Jayanth  R. Banavar$^4$}
\address{$^1$ Institute of Physics, Polish Academy of Sciences,
02-668 Warsaw, Poland}
\address{$^2$ Laboratoire de Physique des Mat{\'e}riaux, Universit{\'e}
Henri Poincar{\'e} Nancy I, F-54506 Vand{\oe}uvre, France} 
\address{$^3$ Center for Polymer Studies, Department of Physics,
Boston University, Boston, MA 02215, USA}
\address{$^4$ Department of Physics and Center for Materials Physics,
104 Davey Laboratory, The Pennsylvania State University, 
University Park, PA 16802}

\address{
\centering{
\medskip\em
{}~\\
\begin{minipage}{14cm}
The master equation for 12-monomer lattice
heteropolymers is solved numerically and the time evolution of the
occupancy of the native state is determined.
At low temperatures,
the median folding time follows the Arrhenius law and
is governed by the longest relaxation time.
For good folders, 
significant kinetic traps 
appear in the folding funnel whereas for bad folders, the traps also occur 
in non-native energy valleys.
{}~\\
{}~\\
{\noindent PACS numbers: 87.15.By, 87.10.+e}
\end{minipage}
}}

\maketitle

The key problem in protein folding is one of dynamics.
Tremendous progress has been made in understanding  equilibrium properties
of simplified lattice models\cite{1}. Such studies have demonstrated the
requirement of thermodynamic stability\cite{1,2}, stability against 
mutations \cite{3}
and a linkage between rapid folding and stability
of the native state\cite{2} ({\sc nat}). Monte Carlo studies of folding
have been helpful in elucidating the folding funnel\cite{5,6}
and meaningful relationships with experiment are being established\cite{7}.
So far, the approaches to studies of the folding dynamics
have been restricted to Monte Carlo simulations that start from a few
randomly chosen initial conformations\cite{8} and the enumeration of 
transition rates between classes of 
conformations which have the same number of contacts and are
a given number of kinetic steps away from the {\sc nat}\cite{9}.
The approximations involved in these approaches remain necessarily untested.

In this letter, we present an exact method to study the dynamics of
short model proteins based on the master equation\cite{10}. 
To illustrate the method,
we present results for sequences made of 12 monomers and placed on a
square lattice. We focus on two sequences, A and B, which 
have good and bad folding properties respectively.
We find that the dynamics of A and B
are superficially similar:  for both, the median 
folding time, $t_{\rm fold}$, and
the longest relaxation time, $\tau _1$ diverge  at low $T$ according
to an Arrhenius law. 
However, what distinguishes the
two cases is the location of the folding transition 
temperature, $T_f$, with respect to the 
temperature $T_{\rm min}$ at which folding to the native state
proceeds the fastest. 
$T_f$ is defined as the temperature at which the equilibrium
value of the probability to occupy the {\sc nat}, $P_0$, crosses 
$\frac{1}{2}$  and is a measure of thermodynamic stability. For bad folders,
$T_f$ is well below $T_{\rm min}$ and thus a substantial occupation
probability for  
the {\sc nat} is found only in a temperature range in which the
dynamics are glassy. 

A deeper understanding of the differences between
A and B is obtained by the identification of  kinetic traps through an
analysis of the eigenvectors corresponding to the longest relaxation
time. The most potent kinetic trap for sequence A is within the
folding funnel and is a few steps away from the {\sc nat}.
The energy needed to exit the trap determines the barrier, $\delta E$,
in the Arrhenius law, $t_{\rm fold} \sim \exp (\delta E /T)$.
For the bad folder, the relevant trap forms its own energy valley 
and exiting it requires
full unfolding. The are many ways to unfold and the effective
$\delta E$ is entropy influenced -- the bottleneck arises from
a search process.

\noindent
{\bf Method:$\;\;$}
Consider a lattice polymer which can exist in $\cal N$ conformations
($\cal N$=15037 for 12-monomer sequences).  
Let $P_{\alpha} = P_{\alpha}(t)$ be the probability of finding the 
sequence in conformation $\alpha$ at time $t$. The master equation is
\begin{equation}
\frac{\partial P_{\alpha}}{\partial t} = \sum_{\beta \neq \alpha}
\left[ w(\beta \to \alpha) P_{\beta} - w(\alpha \to \beta) P_{\alpha} \right]
\end{equation}
where $w_{\alpha \beta}=w(\beta \to \alpha)$ is the transition rate 
from conformation
$\beta$ to conformation $\alpha$. We bring this into a matrix form by
letting $\vec{P} = (P_1,\ldots,P_{\cal N})$ and 
\begin{equation} \label{StochBed}
h_{\alpha \beta } = - w_{\alpha \beta} \leq 0 \;\;
 \mbox{\rm if $\alpha \neq \beta$} \;\; , \;\;
h_{\alpha \alpha} = \sum_{\beta \neq \alpha} w_{\beta \alpha} \;\; .
\end{equation}
The master equation then takes the form of an imaginary-time
Schr\"odinger equation $\partial_t \vec{P} = - \hat{H} \vec{P}$, where
the $h_{\alpha \beta}$ are the matrix elements of $\hat{H}$. 
While this reformulation
is standard \cite{vanK81}, it has regained interest recently because
$\hat{H}$ can often be related to integrable quantum systems \cite{Alca94}.

It is well known \cite{vanK81} 
that the conditions (\ref{StochBed}) are necessary and
sufficient for a matrix $(\hat{H})_{\alpha \beta}=h_{\alpha \beta}$ 
to give rise to a stochastic Markov process. In particular, it follows that
if initially $0\leq P_\alpha \leq 1$ for all conformations, this will hold
true at all subsequent times. Time-dependent averages for any observable
$X$ are found from
\begin{equation}
\langle X \rangle (t) = \langle s| \hat{X} e^{-\hat{H}t} | P_{in}\rangle
\end{equation}
where $\hat{X}$ is the matrix representation of $X$, $\langle s| =
(1,\ldots,1)$ is the left steady state of $\hat{H}$ with
$\langle s| \hat{H}=0$ and $|P_{in}\rangle = 
\sum_{\alpha} P_{\alpha}(0) |\alpha \rangle$ is the
initial state. The spectrum of relaxation times 
$\tau_{\alpha} = 1/\mbox{\rm Re } E_{\alpha} \geq 0$
follows directly from the eigenvalues $E_{\alpha}$ of $\hat{H}$. 

An important special case arises if the right steady state $|s\rangle 
=\sum_{\alpha} P_{\alpha}^{\rm eq} |{\alpha}\rangle$ 
is related to a Hamiltonian
${\cal H}$ through $P_{\alpha}^{\rm eq} \sim e^{-{\cal H}_{\alpha}/T}$. 
This happens provided
the detailed balance condition
\begin{equation} \label{DetGG}
w_{\alpha \beta} P_{\beta}^{\rm eq} = 
w_{\beta \alpha} P_{\alpha}^{\rm eq}
\end{equation}
is satisfied and then $P_{\alpha}^{\rm eq}$ is indeed a steady-state 
solution of the
master equation. Eq. (\ref{DetGG}) is satisfied by
$w_{\alpha \beta}=f_{\alpha \beta}
 \exp \left[ -({\cal H}_{\alpha} - {\cal H}_{\beta}) /2T \right]$ 
provided $f_{\alpha \beta}=f_{\beta \alpha}$.
Here, we choose $w_{\alpha \beta}=w_{\alpha \beta}^{(1)}+
w_{\alpha \beta}^{(2)}$, where 
\begin{equation}
w_{\alpha \beta}^{(\sigma)} = \frac{1}{\tau _0} R_{\sigma} \left( 1 + \exp
\left( \frac{ {\cal H}_{\alpha} - {\cal H}_{\beta} }{T} 
\right) \right)^{-1} 
\end{equation}
with $R_1 + R_2 =1$. Here, $\sigma$ refers to 
the single- and double-monomer moves and $\tau _0$ is a microscopic 
time scale.
It is understood that $w_{\alpha \beta}^{(\sigma)}=0$ if there is no move of
type $\sigma$ linking $\beta$ with $\alpha$. 
This choice guarantees that transition
rates are finite and bounded for all temperatures. In analogy to
ref.\cite{2}, we focus on $R_1$=0.2 and take the single and two-monomer
(crankshaft) moves as in ref.\cite{9}.

Because of the detailed balance condition, 
the eigenvalues $E_{\alpha}$ are not calculated by
diagonalizing $\hat{H}$ directly, but 
by diagonalizing an auxiliary matrix $\hat{K}$ with
elements
\begin{equation}
k_{\alpha \beta} = \frac{v_{\beta}}{v_{\alpha}} h_{\alpha \beta} =
k_{\beta \alpha} \;\;,
\end{equation}
where $v_{\alpha}= e^{-{\cal H}_{\alpha}/(2T)}$. 
The right eigenvectors $|E_{\alpha}\rangle$ of
$\hat{H}$ are found from the eigenstates $\hat{K} |F_{\alpha}\rangle
= E_{\alpha} |F_{\alpha}\rangle$ via $|E_{\alpha}\rangle =
 v_{\alpha} |F_{\alpha}\rangle$, but must still
be normalized for a probabilistic interpretation. It follows that the
eigenvalues $E_{\alpha}$ are real and positive and that the eigenstates 
span a complete basis \cite{vanK81}. The eigenvector
corresponding to $E_{\alpha}=0$, i.e. to the infinite 
relaxation time, determines the equilibrium occupancies of the
conformations. The longest finite relaxation time 
$\tau _1 =1/E_1$ is found from the smallest non-zero eigenvalue $E_1$.

The practical calculation of the eigenvalues $E_{\alpha}$ 
and of the lowest
two eigenvectors of $\hat{H}$ is done through the standard symmetric
Lanczos algorithm without reorthogonalization. Since memory requirements
are the essential limit of the method, it is important that besides the
matrix elements, only two more vectors have to be kept in memory
\cite{Chri93}. To find the eigenvectors, we follow the suggestion of
Dagotto \cite{Dago94} to run the Lanczos algorithm {\em twice}. In the first 
pass, we find the eigenvalues and the similarity transformation which
diagonalizes the intermediate tridiagonal matrix constructed from
$\hat{K}$. In the second pass, this information is used to accumulate
the eigenvectors
from the intermediate vectors which in this way need
not be kept in memory. 
The number of Lanczos iterations needed to achieve good
convergence varied between 200 and 2500, depending on the temperature 
and also the sequence considered (for sequence A, the convergence is
more rapid than for B).
The time-dependence state vector $\vec{P}(t)$ at time $t=n\tau _0$ can
be obtained by applying $n$ times the recursion $\vec{P}((n+1)\tau _0)=
(1 + \hat{H}) \vec{P}(n\tau _0)$ to $\vec{P}_{in}$.

\noindent
{\bf Results:$\;\;$}
The energies of a sequence are determined by the Hamiltonian
${\cal H} = \sum_{ij} B_{ij}  \Delta _{ij} $,
where the contact interaction, $B_{ij}$, is assigned to monomers
$i$ and $j$ which are geometrical nearest neighbors on the lattice
but are not neighbors along the sequence -- the condition symbolized
by $\Delta _{ij}$. For 12 monomers, there are 25 contact energies
which we pick as Gaussian numbers of unit dispersion and with a mean value
around -1 to provide an overall 
attraction\cite{coupl}. Sequence B has couplings identical in strength
to those in A but the assignment to monomers is permuted\cite{coupl}.
  
The ground state of sequence A is maximally compact and fills the
$3 \times 4$ lattice. 
The probability, that it is occupied at 
time $t$, $P_0(t)$ depends on the initial condition. Figure 1 shows
the evolution of $P_0(t)$ from three different initial states. 
The solid line corresponds to an  initial state 
in which all conformations have  equal probability of $1/{\cal N}$ of
being occupied.
The broken line is for the situation in which the system is initially in the 
{\sc nat}. Finally, the dotted line corresponds to the initial
state being the kinetic ``trap'' conformation \cite{15}
which is the strongest obstacle in reaching equilibrium.

The trap is determined
by studying the eigenvector corresponding to the longest relaxation 
time and by identifying the local energy minima
which have the largest weights
at low $T$. The largest weight is associated with the {\sc nat}
whereas the second largest corresponds to the most relevant trap.
In the limit of $T \rightarrow 0$,  weights associated with all other
states become insignificant. 

Figure 1 shows that the equilibrium value of $P_0$
is reached in essentially the same time, independent
of the initial state
because the long time dynamics is determined by just one mode with a relaxation
time  $\tau _1$. The time, $t_{\frac{1}{2}}$, 
needed to reach, say, half of the
equilibrium value, does depend on the initial state --
it is significantly longer for the trap state.

The top of Figure 2 shows the {\sc nat} and trap conformations.
The latter is 2.5404 energy units above the {\sc nat}. The overall least costly
path (energetically) between the trap and the  {\sc nat} 
involves at least 10 steps
and requires an increase of 4.5323 above the trap energy. The most costly
step in this trajectory requires an energy of 2.8823 to move
monomer 12 away from monomers 5 and 7.

Figure 2 summarizes results obtained from the master equation, when the initial 
state is of uniform occupancy and compares them to the median folding time
obtained through Monte Carlo simulations  which satisfy detailed
balance conditions along the lines described in ref. \cite{9}. 
The low temperature behavior of $t_{\rm fold}$ follows the Arrhenius law,
and $\delta E$ is close to the energy needed to exit the kinetic trap.
In this region, $t_{\rm fold}$ is
proportional to $\tau _1$. This longest relaxation time essentially
coincides with $t_{\frac{1}{2}}$ when the kinetic trap
is the initial state.

The Arrhenius behavior sets in fairly close to $T_{\rm min}$ where
$t_{\rm fold}$ displays a minimum.  On the high temperature side of
$T_{\rm min}$, the characteristic times related to the approach to equilibrium
no longer have any relationship to $t_{\rm fold}$  and the values of
$P_0$ are small (0.064 at $T$=1.2).
The physical situation changes now: reaching the {\sc nat}
now is controlled by fluctuations in equilibrium
and is governed by the statistics of rare events.

Figure 3 summarizes the dynamical data for sequence B for which $T_f$
is substantially  below $T_{\rm min}$ and signifies bad folding properties.
The {\sc nat} for B is not maximally compact and is doubly
degenerate as shown at the top of Figure 3. The two states differ
merely by placement of one monomer and, when studying folding, are
considered as an effective single state. The overall shape of the 
temperature dependence of $t_{\rm fold}$ is similar to that for sequence 
A and the low $T$ Arrhenius law is also obeyed. The kinetic trap state,
also shown in Figure 3, is very close in shape to the {\sc nat} and
it differs from it only by one contact. This state, however, is very remote
kinetically: all trajectories which lead from the trap to the
{\sc nat} must go through an unfolded state and take at least 31 steps
with the biggest single step energy increase of 2.7478.
This trap is not in the folding funnel of the {\sc nat} -- the energy
landscape is thus very rugged. The $\delta E$ of the Arrhenius law is 
close to 3.55 and is expected to have a substantial entropy
contribution at non-zero temperatures
due to many possible choices of the trajectories.

The method presented in this letter offers ways of studying  kinetic traps
systematically. In particular, existing truncation techniques for the 
diagonalization of large matrices might be fruitfully employed in
extending the method to longer protein chains.

We thank D. Cichocka and O. Collet for discussions.
This work was supported by  KBN (Grant No. 2P03B-025-13),
Polonium, CNRS-UMR 7556, The Fulbright Foundation (JK),
NASA, The Center for Academic Computing 
and the Applied Research Laboratory at Penn State.

\newpage

\noindent
FIGURE CAPTIONS
\begin{description}
\item 1. Probability of occupation of the native state, $P_0(t)$,
of sequence A, 
for three values of the temperatures, indicated on the right.
$P_0(\infty)$ agrees with the equilibrium value.
For sequence B, the values of $P_0(\infty)$ are significantly lower
than that for sequence A -- for example, at $T$=0.6, $P_0(\infty)$=0.0752.
The circles correspond to Monte Carlo results, based on 200 random
starting conformations.

\item 2. Top: The {\sc nat} and 'trap' conformations 
and their energies for sequence A. The enlarged circle shows the first
monomer.
Main: Dynamical data for the folding.
The solid line marked by $t_{\rm fold}$ gives the median folding time
derived from  1000 Monte Carlo trajectories.
The solid line $\tau _1$ is the longest relaxation time.
The dotted line $t_{eq}$ is the time to reach equilibrium  
from the initial state of uniform occpancy. 
The broken line $t_{\frac{1}{2}}$ with the black circles  gives the
time to reach $\frac{1}{2} P_0$ 
from this initial state. The
open circles indicate the same but the trap is taken as initial state.
The dotted line $t_A$ is a fit of the Monte Carlo data
to the Arrhenius law with $\delta E$=2.76. The arrow at the
top indicates the value of the folding temperature.

\item 3. Same as Figure 2, but for sequence B. 
For the curve $t_A$, $\delta E$=3.55. There are two {\sc nat} conformations
with the same energy.

\end{description}

\begin{thebibliography}{99} \vspace{-15mm}

\bibitem{1} 
P. G. Wolynes, J. N. Onuchic, and D. Thirumalai,
Science {\bf 267}, 1619 (1995);
K. A. Dill, S. Bromberg, S. Yue, K. Fiebig, 
K. M. Yee, D. P. Thomas, and H. S. Chan, 
Protein Sci. {\bf 4}, 561 (1995);
C. J. Camacho and D. Thirumalai, Proc. Nat. Acad. Sci. USA {\bf 90},
6369 (1993); 
H. S. Chan and K. A. Dill, Phys. Today {\bf 46}, 24 (1993).

\bibitem{2}
A. Sali, E. Shakhnovich, and M. Karplus, Nature {\bf 369}, 248 (1994).

\bibitem{3}
H. Li, R. Helling, C. Tang, and N. Wingreen, Science {\bf 273}, 666 (1996);
M. Vendruscolo, A. Maritan, J. R. Banavar, Phys. Rev. Lett. {\bf 78},
3967 (1997). 


\bibitem{5} P. E. Leopold, M. Montal, and J. N. Onuchic, Proc. Natl.
Acad. Sci. USA {\bf 89}, 8721 (1992).

\bibitem{6} M. Cieplak, S. Vishveshwara, and J. R. Banavar, Phys. Rev.
Lett. {\bf 77}, 3681 (1996); M. Cieplak and J. R. Banavar, Folding \&
Design {\bf 2}, 235 (1997).


\bibitem{7} J. N. Onuchic, P. G. Wolynes, Z. Luthey-Schulten,
Proc. Natl. Acad. Sci. {\bf 92}, 3626 (1995); J. D. Bryngelson,
J. N. Onuchic, N. D. Socci, and P. G. Wolynes, Proteins: Struct. Funct.
and Genet. {\bf 21}, 167 (1995).

\bibitem{8} see, e.g., N. D. Socci and J. N. Onuchic, J. Chem. Phys.
{\bf 101}, 1519 (1994).

\bibitem{9} H. S. Chan and K. A. Dill, J. Chem. Phys. {\bf 99}, 2116 (1993);
J. Chem. Phys. {\bf 100}, 9238 (1994).

\bibitem{10}This approach has been motivated by similar studies
of frustrated Ising spin clusters in J. R. Banavar,
M. Cieplak, and M. Muthukumar, J. Phys. C {\bf 18}, L157 (1993).

\bibitem{vanK81} 
N.G. van Kampen, {\it Stochastic Processes in Physics and
Chemistry}, North Holland (Amsterdam 1981); 
J. Schnakenberg, Rev. Mod. Phys. {\bf 48}, 571 (1976).

\bibitem{Alca94} F.C. Alcaraz, M. Droz, M. Henkel and V. Rittenberg,
Ann. of Phys. {\bf 230}, 250 (1994).

\bibitem{Chri93} P. Christe and M. Henkel, {\it Introduction to Conformal
Invariance and its Applications to Critical Phenomena}, Springer
(Heidelberg 1993), ch. 9.

\bibitem{Dago94} E. Dagotto, Rev. Mod. Phys.  {\bf 66}, 763 (1994).

\bibitem{coupl}
The couplings $B_{i,j}$ are 
$B_{1,4}=-0.6308$, $B_{1,6}=-2.0474$, $B_{1,8}=-0.7504$, 
$B_{1,10}=-1.3210,
B_{1,12}= -0.5289, B_{2,5}=-2.3830, B_{2,7}=-1.4923,
B_{2,9}=-0.1592, B_{2,11}=-1.2074, 
B_{3,6}=-1.1705, B_{3,8}=0.1223,
B_{3,10}= -0.8999, B_{3,12}=-0.4610, B_{4,7}=-0.4581,
B_{4,9}=-1.9629,  B_{4,11}=-1.5981,
B_{5,8}=-1.5677, B_{5,10}=-0.8795,
B_{5,12}= -0.9902, B_{6,9}=0.2053, B_{6,11}=-1.2078,
B_{7,10}=-0.3809, B_{7,12}=-1.8921, B_{8,11}=-1.6500, B_{9,12}=-0.0989$ 
for sequence A. 
For sequence B, 7 of the couplings are rearranged: contact (1,10) is  
interchanged with (5,10), (2,5) with (2,9), (3,8) with (6,11), 
(4,7) is assigned to (6,9), (7,12) to (4,7), and (4,7) to (6,9).

\bibitem{15}
L. A. Mirny, V. Abkevich, and E. I. Shakhnovich, Folding \& Design
{\bf 20}, 103 (1996).

\end{thebibliography}
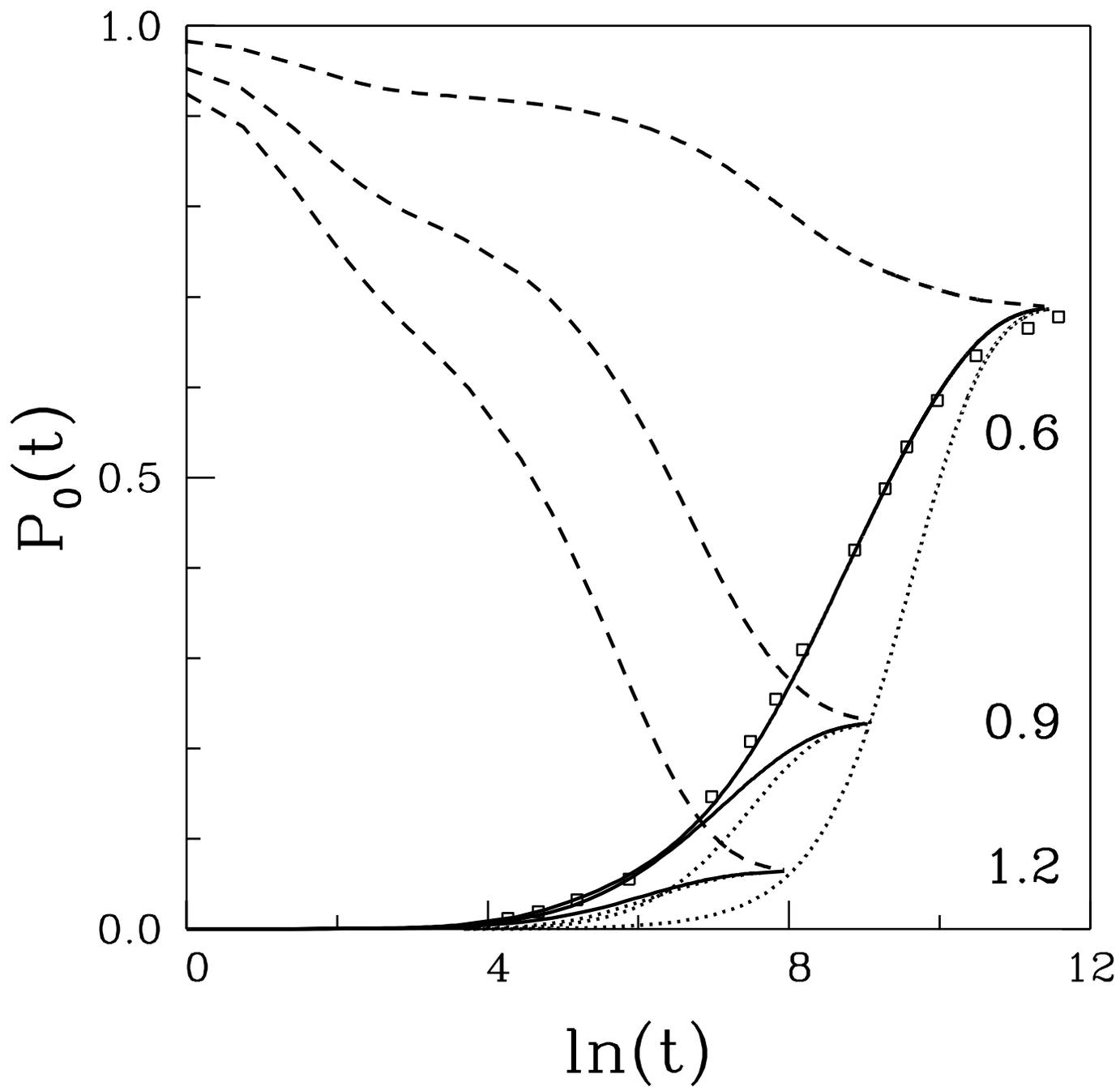
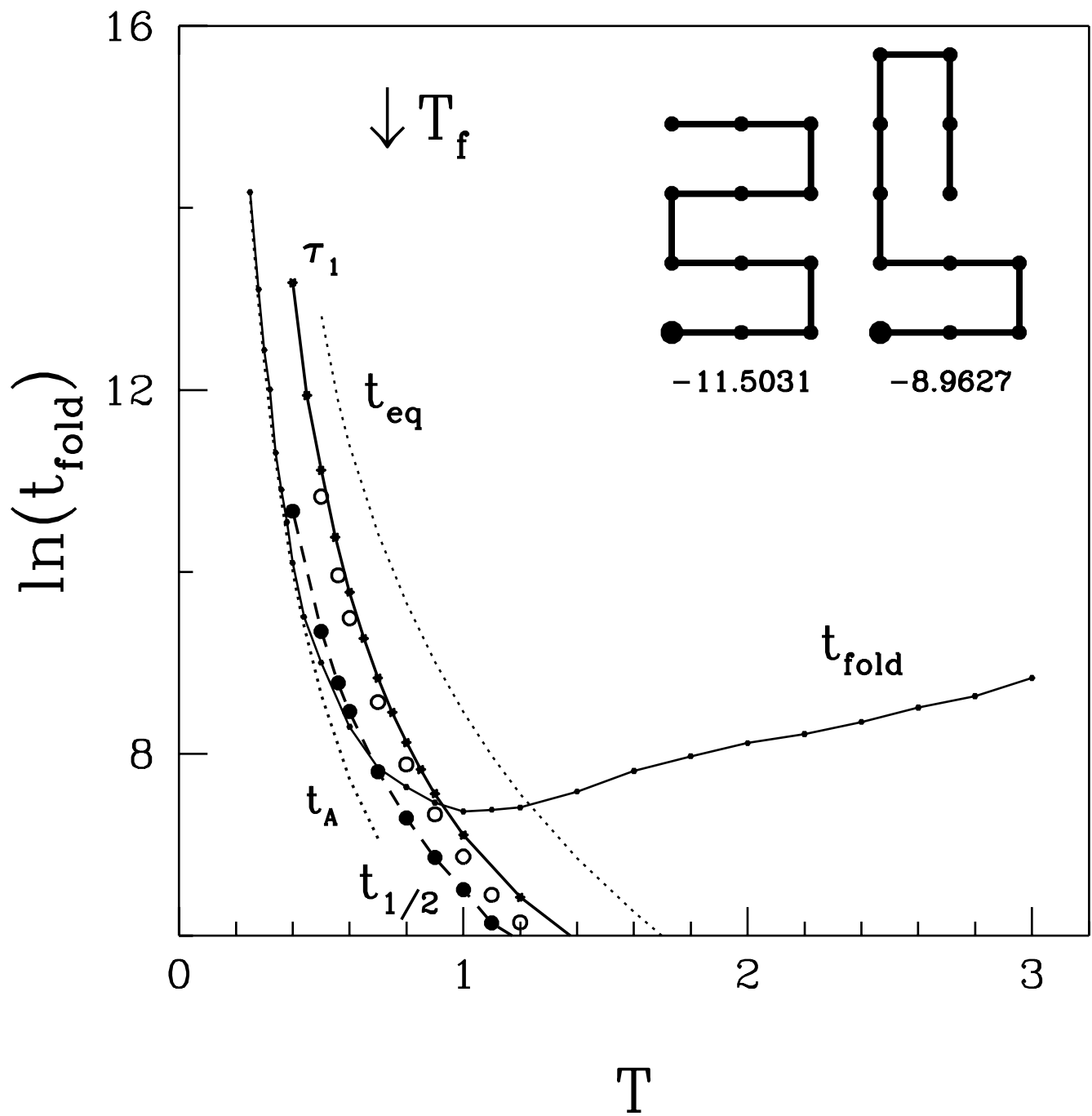
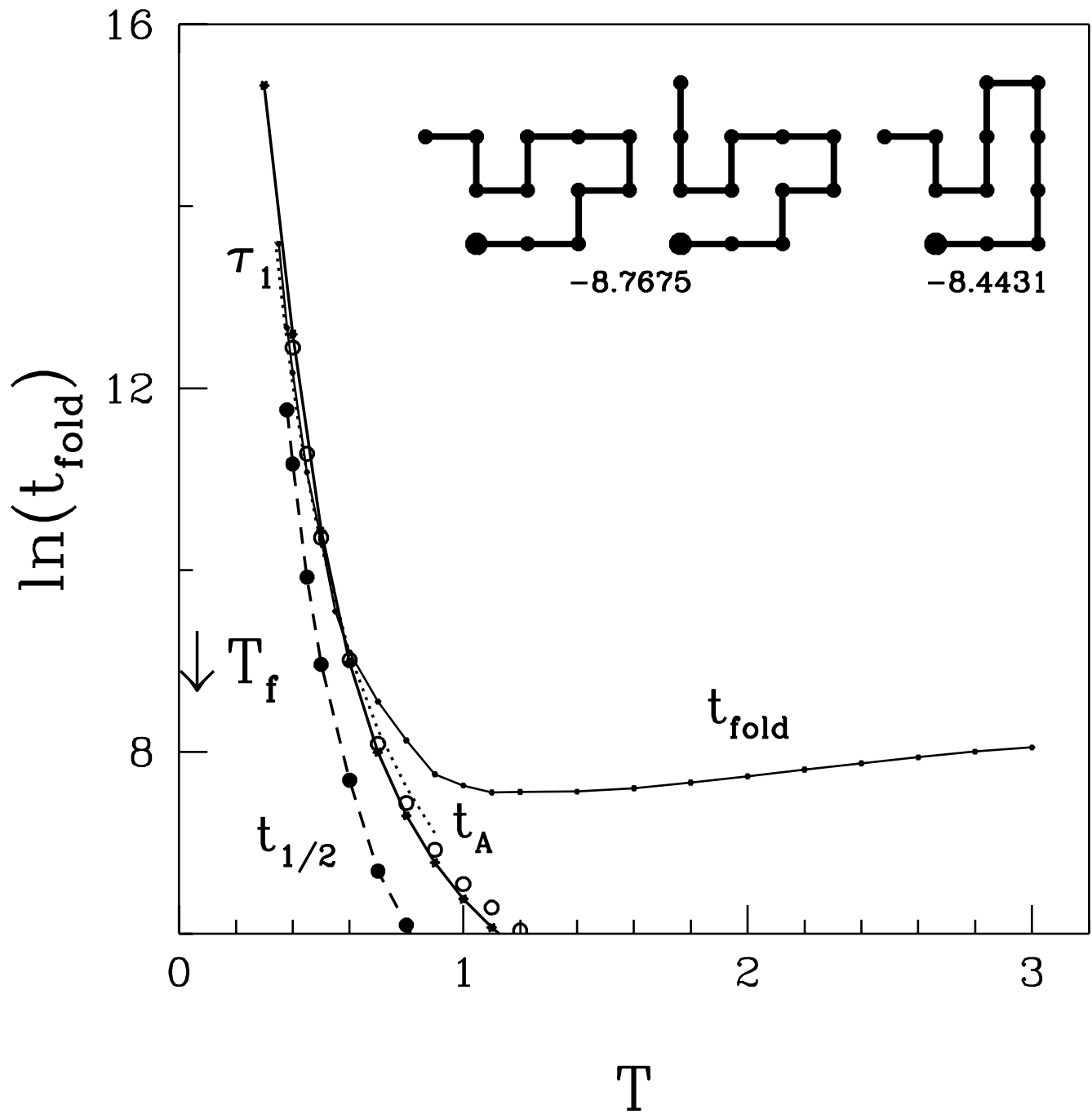
\end{document}